\documentclass[a4paper]{emulateapj}
\usepackage{graphicx}
\usepackage{natbib}
\usepackage{epsf}

\def\solmas{$\mathrm{M_\odot}$~}
\def\solmasp{$\mathrm{M_\odot}$}

\def\numdensu{cm$^{-3}$}

\def\apjl{{\em ApJ}}
\def\apjs{{\em ApJS}}
\def\aap{{\em A\&A}}

\def\simless{\mathbin{\lower 3pt\hbox
   {$\rlap{\raise 5pt\hbox{$\char'074$}}\mathchar"7218$}}}
\def\simgreat{\mathbin{\lower 3pt\hbox
   {$\rlap{\raise 5pt\hbox{$\char'076$}}\mathchar"7218$}}}
\newcommand{\zfrag}{${\rm Z} = 10^{-5} \: {\rm Z_{\odot}} $}
\newcommand{\zsix}{${\rm Z} = 10^{-6} \: {\rm Z_{\odot}} $}
\newcommand{\zleqsix}{${\rm Z} \leq 10^{-6} \: {\rm Z_{\odot}} $}

\begin{document}

\title{The First Stellar Cluster}

\author
{Paul C.\ Clark$^1$, Simon C.\ O.\ Glover$^2$, Ralf S.\ Klessen$^1$}

 \affil{$^{1}$Zentrum f\"ur Astronomie der Universit\"at Heidelberg, Institut
f\"ur Theoretische  Astrophysik, Albert-Ueberle-Str.\ 2, 69120 Heidelberg, Germany
\break email: pcc@ita.uni-heidelberg.de, rklessen@ita.uni-heidelberg.de \\
$^2$Astrophysikalisches Institut Potsdam, An der Sternwarte 16, D-14482 Potsdam,
Germany  \break email: sglover@aip.de}

\begin{abstract} 

We report results from numerical simulations of star formation in the early universe that
focus on gas at very high densities and very low metallicities. We argue that the  gas in
the central regions of protogalactic halos will fragment as long as it carries sufficient
angular momentum. Rotation leads to the build-up of massive disk-like structures which
fragment to form protostars. At metallicities ${\rm Z} \approx 10^{-5}\,{\rm Z}_{\odot}$,
dust cooling  becomes effective and leads to a sudden drop of temperature at densities above
$n = 10^{12}\,$cm$^{-3}$. This induces vigorous fragmentation, leading to a very
densely-packed cluster of low-mass stars. This is the first stellar cluster. The mass
function of stars peaks below $1\,$M$_{\odot}$, similar to what is found in the solar
neighborhood, and comparable  to the masses of the very-low metallicity subgiant stars
recently discovered in the  halo of our Milky Way. We find that even purely primordial gas
can fragment at  densities $10^{14}\,$cm$^{-3} \le n \le 10^{16}\,$cm$^{-3}$, although the
resulting mass function contains only a few objects (at least a factor of ten less than the
\zfrag~mass function), and is biased towards higher masses. A similar result is found for
gas with \zsix. Gas with Z $\le 10^{-6}\,$Z$_{\odot}$ behaves roughly isothermally at these
densities  (with polytropic exponent $\gamma \approx 1.06$) and the massive disk-like
structures  that form due to angular momentum conservation will be marginally unstable. As 
fragmentation is less efficient, we expect stars with Z $\le 10^{-6}\,$Z$_{\odot}$  to be
massive, with masses in excess of several tens of solar masses, consistent with the results
from previous studies.

\end{abstract}

\maketitle

\section{Introduction} Detailed numerical simulations of the formation of the first stars,
the  so-called population III stars, indicate that they were probably massive, with masses
greater than $20 \: {\rm M_{\odot}}$ \citep{abn02,bcl02,yoha06,oshn07}. The fact that no
population III star has ever been observed in the Milky Way provides some observational
support for this prediction \citep{tum06}. However, a number of low-mass, extremely
metal-poor, stars with $[{\rm Fe}/{\rm H}] < -3$ have been discovered in the  Galactic halo
\citep{bc05}, suggesting that the distribution of stellar masses is sensitive to even very
low levels of metal enrichment.  Explanations of this apparent change in the IMF have
concentrated on the fact that metal-enriched gas has more coolants than its primordial
counterpart.  These coolants are suggested to provide an opportunity for efficient
fragmentation, since they can keep the gas temperature lower during the collapse process
than is possible with pure H$_{2}$ cooling in the primordial gas. If this is the case, then
the final IMF will likely contain at least some low-mass stars, even if it still differs
significantly from the present-day local IMF.

If metal enrichment is the key to the formation of low-mass stars, then logically
there  must be some critical metallicity ${\rm Z_{\rm crit}}$ at which the formation
of low mass stars first becomes possible. However, the value of  ${\rm Z_{\rm crit}}$
is a matter of ongoing debate. Some models suggest that low mass star formation
becomes possible only once atomic fine-structure line cooling from carbon and oxygen
becomes effective \citep{bfcl01,bl03,san06,fjb07}, setting a value for ${\rm Z_{\rm
crit}}$ at around $10^{-3.5} \: {\rm Z_{\odot}}$. Another possibility, and the one
that we explore with this paper, is that low mass star formation is a result of
dust-induced fragmentation occurring at high densities, and thus at a very late stage
in the protostellar collapse \citep{sch02,om05,sch06}. In this model,
$10^{-6}~\simless~{\rm Z_{\rm crit}}~\simless~10^{-4}~{\rm Z_{\odot}}$, where much of
the uncertainty in the predicted value results from uncertainties in the dust
composition and the degree of gas-phase depletion \citep{sch02,sch06}.

The recent simulations performed by \citet{to06}, which model the collapse of very high
density protogalactic gas, provide some support for the dust-induced fragmentation model.
Using a simple piecewise polytropic equation of state to describe the thermal evolution of
extremely metal-poor protogalactic gas, \citet{to06} show that fragmentation can occur at
metallicities as low as ${\rm Z} = 10^{-6} \: {\rm Z_{\odot}}$, and that it becomes more
effective as the metallicity increases. However, their study considered only the limiting
case of gas with zero angular momentum. Owing to the absence of angular  momentum, the
fragments formed in their simulation do not survive for longer than a  dynamical time, as
they simply fall to the center of the potential well, where they merge with other fragments.
It is also unclear whether the fragmentation process would be as effective if the angular
momentum of the gas were non-zero, as would be expected in reality.

We present the results of simulations of the high-density, dust-cooling dominated
regime that improve on those of \citet{to06} by including the effects of rotation,
and by following a much larger dynamical range of the collapse (ten orders of
magnitude in density), as well as employing an equation of state (EOS) which follows
that of Omukai et al.\ (2005) more closely. We also perform some simulations of purely
primordial gas for comparison. A key feature is the use  of sink particles (Bate,
Bonnell \& Price 1995) to capture the formation and evolution of multiple collapsing
cores, which enables us to follow the evolution of the star-forming gas over several
free-fall timescales and thus to  model the build-up  of a stellar cluster. This
differs from previous calculations which either follow the collapse of a single core
to high densities \citep{abn02,bl04,yoha06}, or use sink particles to capture low 
density ($n < 10^{6}$ \numdensu) fragmentation \citep[e.g.][]{bcl02}.

In following section, we give details of the numerical model (\S\ref{code}) and the initial
set-up of the simulations (\S\ref{setup}). The main results of our study are presented in
\S\ref{results} and we discuss the origin of the fragmentation in \S\ref{discussion}.
Potential caveats with the current model are highlighted in \S\ref{caveats}, and there is a
summary of our findings in \S\ref{conclusions}.

\begin{figure}[t]
\includegraphics[width=3.1in,height=1.75in]{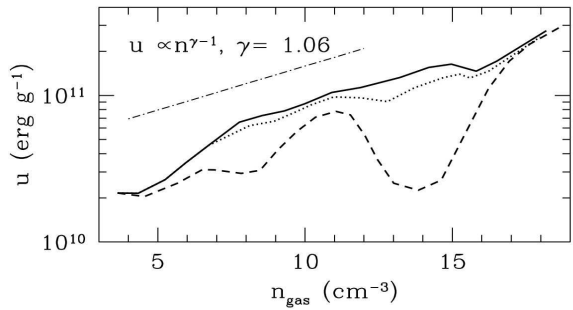}

\caption{\label{fig:EOS} The three equations of state (EOSs) from Omukai et~al. (2005) that are used
in our study. The primordial case (solid line), \zsix~(dotted line), and
\zfrag~(dashed line), are shown alongside an example of a polytropic EOS with an
effective $\gamma = 1.06$. }

\end{figure}

\begin{figure*}[t]
\centerline{
\includegraphics[width=6.4in,height=4.18in]{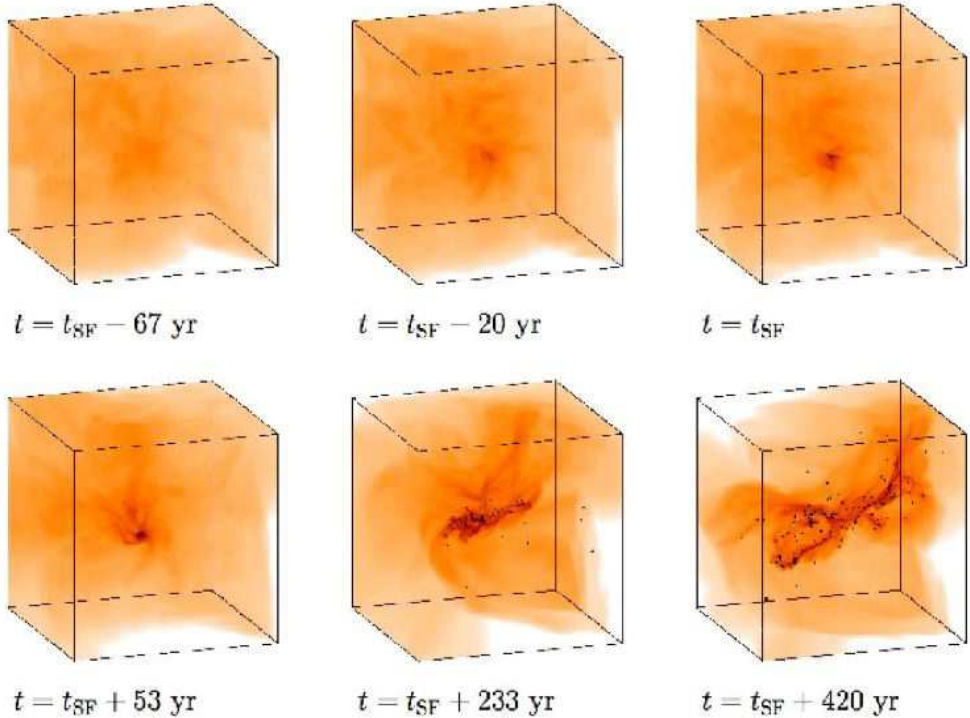}
}

\caption{\label{fig:sequence} Time evolution of the density distribution in the innermost 400 AU of the
protogalactic halo shortly before and shortly after the formation of the first protostar at
$t_{\rm SF}$. We plot only gas at densities above $10^{10}\,$cm$^{-3}$.  The dynamical
timescale at a density $n = 10^{13}\,$cm$^{-3}$ is of the order of only 10 years.  Dark dots
indicate the location of protostars as identified by sink particles forming at $n \ge
10^{17}\,$cm$^{-3}$. Note that without usage of sink particles we would not have been able
to follow the build-up of the protostellar cluster beyond the formation of the first object.
There are 177 protostars when we stop the calculation at $t = t_{\rm SF} + 420\,$yr. They
occupy a region roughly a hundredth of the size of the initial cloud. With
$18.7\,$M$_{\odot}$  accreted at this stage, the stellar density is $2.25 \times
10^{9}\,$M$_{\odot}\,$pc$^{-3}$.}

\end{figure*}

\begin{figure*}[t]
\begin{center}
\unitlength1cm
\begin{picture}(18,5.5)
\put(1.0, 0.50){\includegraphics[width=1.8in,height=1.8in]{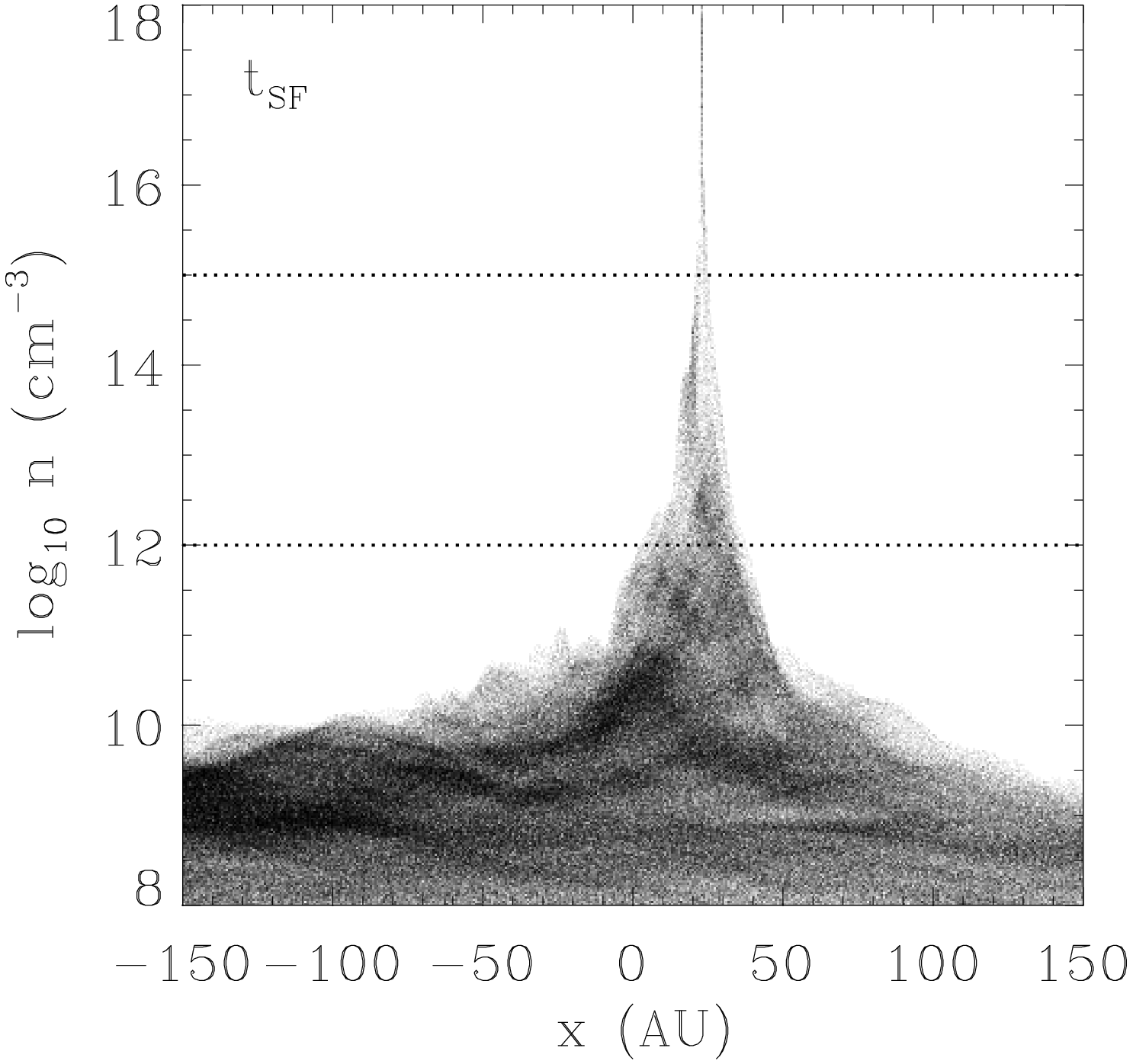}}
\put(7.0, 0.50){\includegraphics[width=1.8in,height=1.8in]{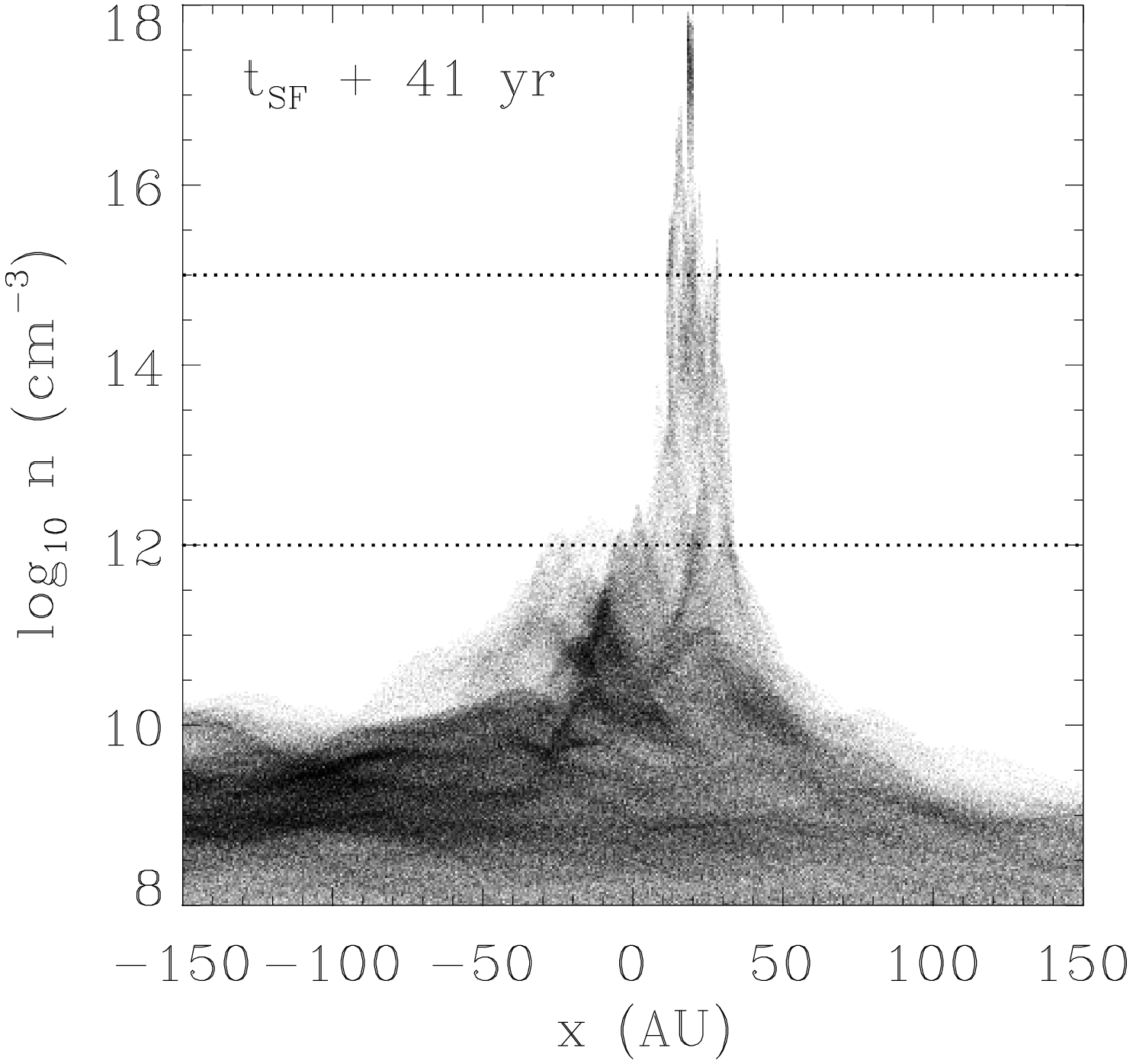}}
\put(13.0, 0.50){\includegraphics[width=1.8in,height=1.8in]{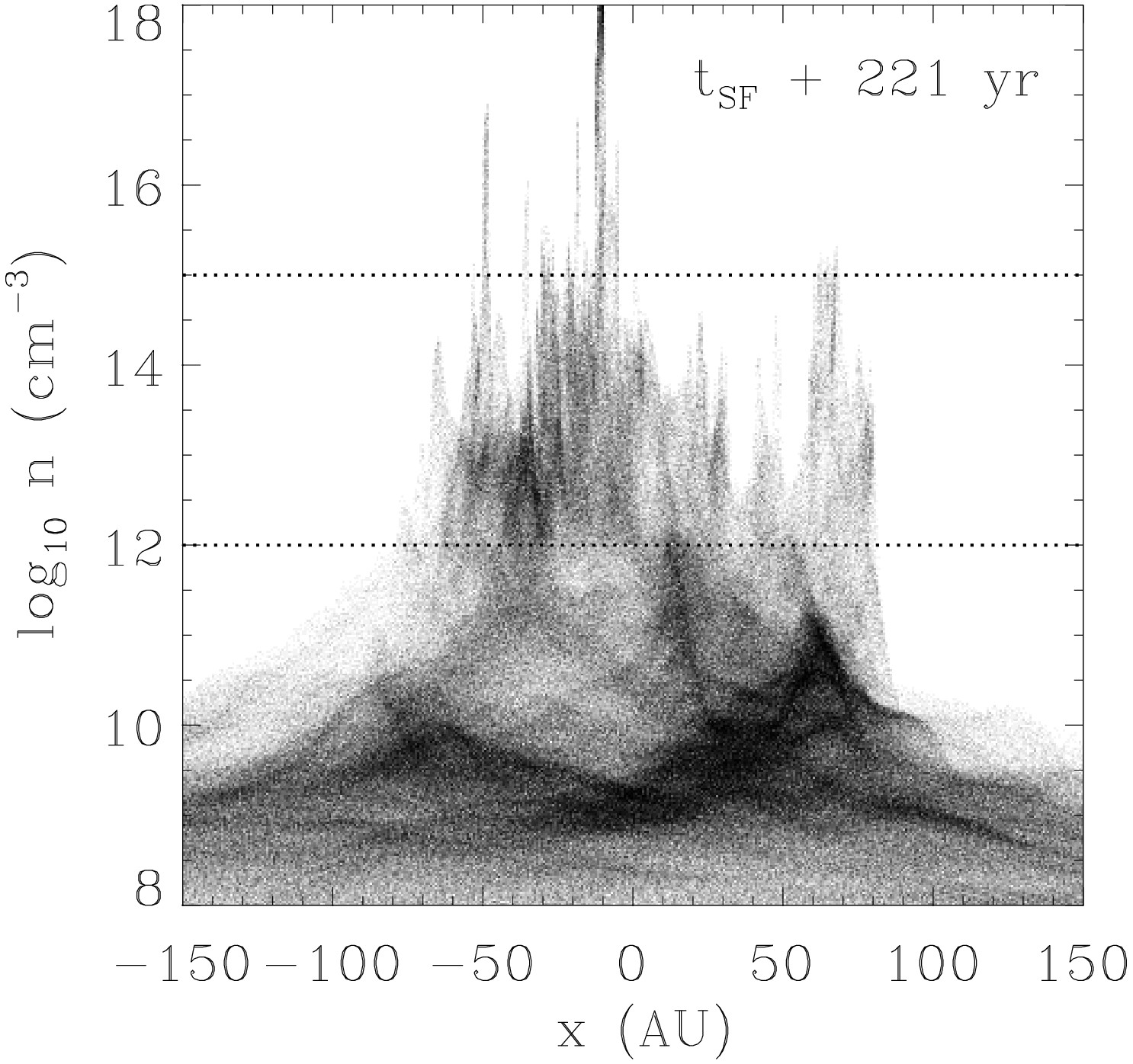}}
\end{picture}
\end{center}
\caption{ \label{fig:density} We illustrate the onset of the fragmentation process in the high resolution
\zfrag~simulation. The graphs show the densities of the particles, plotted as a
function of their x-position. Note that for each plot, the particle data has been centered on
the region of interest. We show here results at three different output times, ranging from
the time that the first star forms ($t_{\rm sf}$) to 221 years afterwards. The densities 
lying between the two horizontal dashed lines denote the range over which
dust cooling lowers the gas temperature.}

\end{figure*}

\begin{figure*}[t]
\centerline{
	\includegraphics[width=2.1in,height=2.1in]{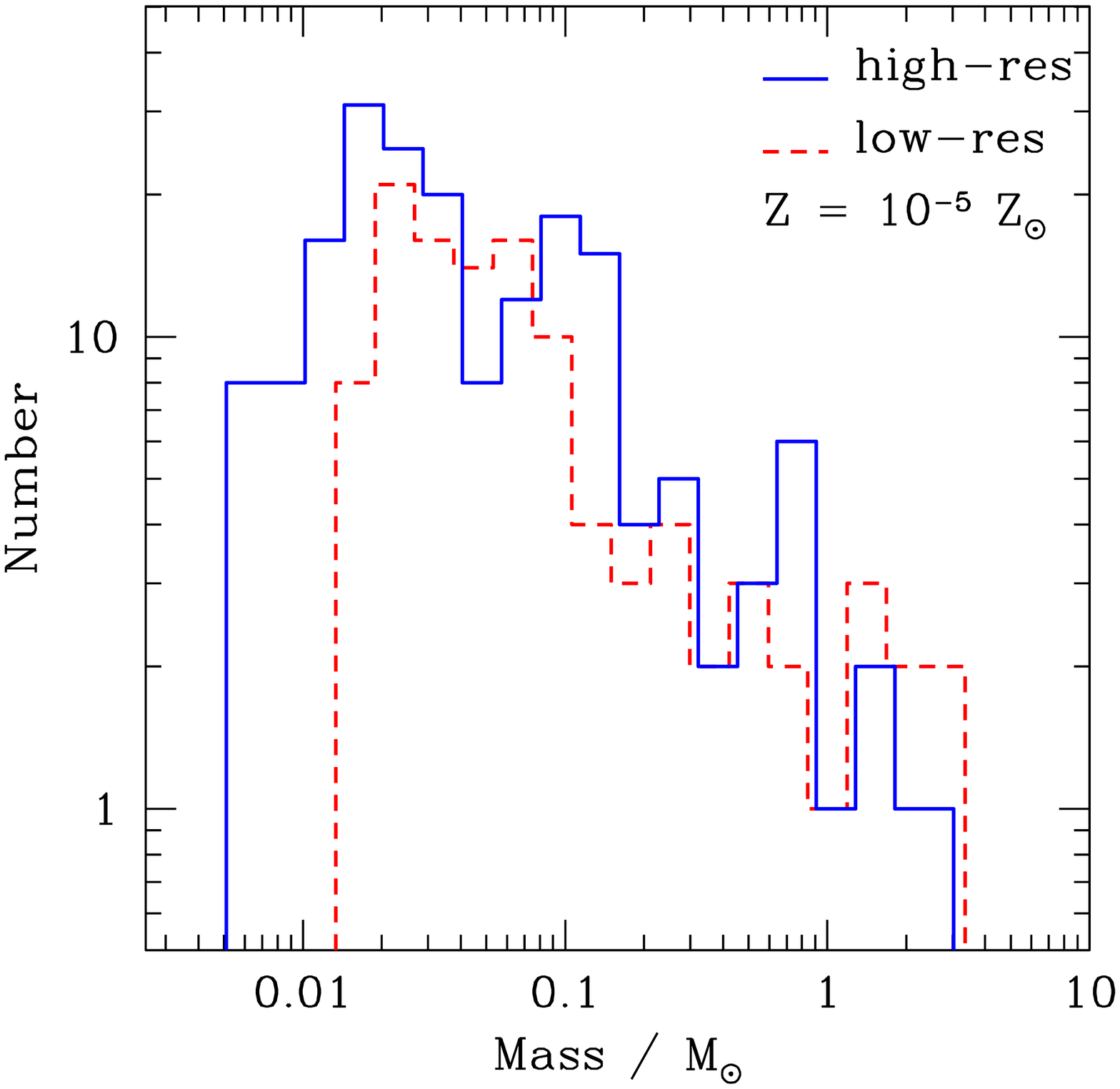}
	\includegraphics[width=2.1in,height=2.1in]{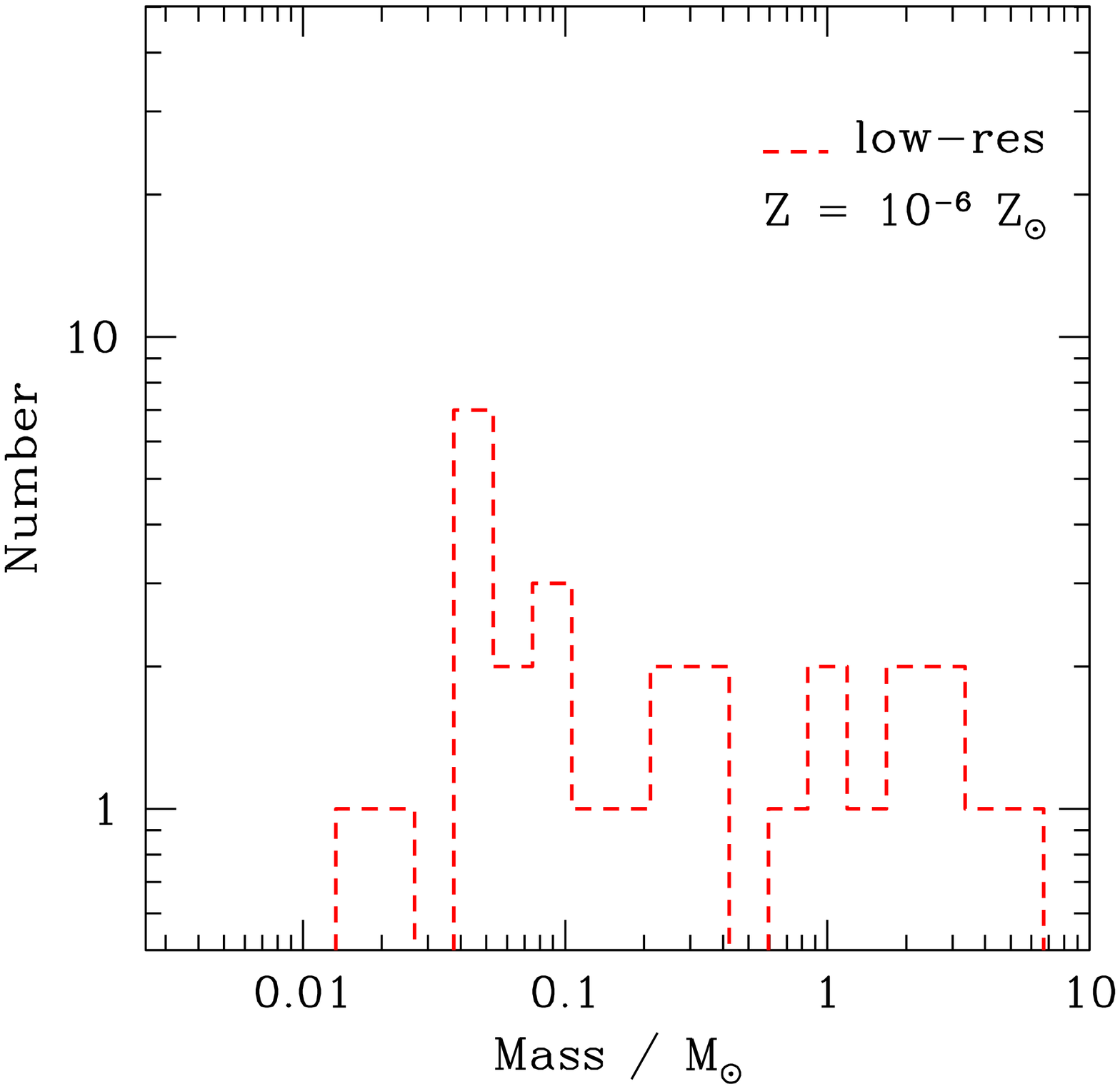}
	\includegraphics[width=2.1in,height=2.1in]{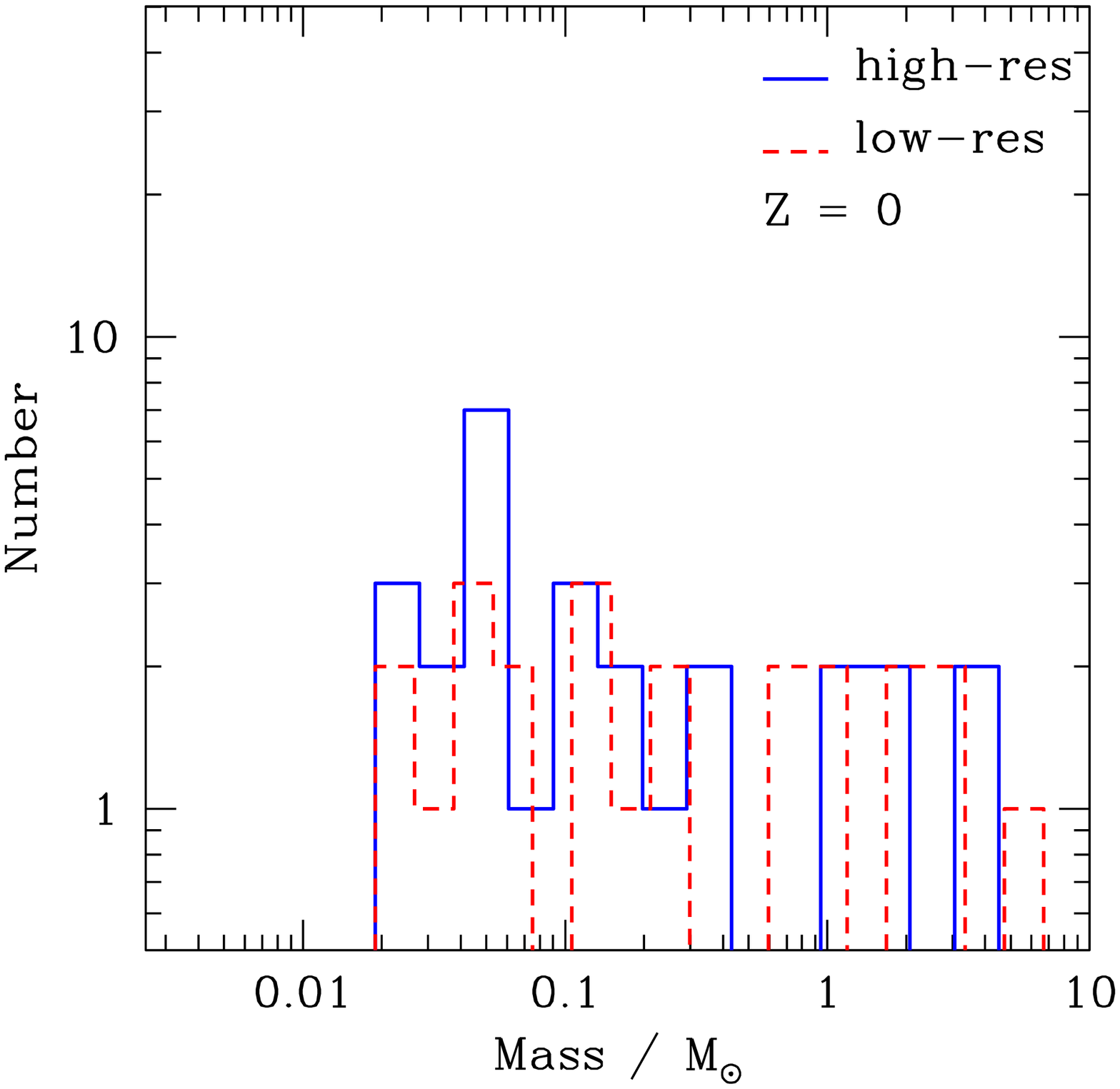}
}

\caption{\label{fig:masses}  Mass functions resulting from simulations with metallicities
${\rm Z} = 10^{-5} \: {\rm Z_{\odot}}$ (left-hand panel), ${\rm Z} = 10^{-6} \: {\rm
Z_{\odot}}$ (center panel), and ${\rm Z} = 0$ (right-hand panel). The plots refer to the
point in each simulation at which 19 \solmas of material has been accreted (which occurs at
a slightly different time in each simulation). The mass resolutions are 0.002 \solmas and
0.025 \solmas for the high and low resolution simulations, respectively.  Note the
similarity between the results of the low-resolution and high-resolution simulations. The
onset of dust-cooling in the \zfrag~cloud results in a stellar cluster which has a mass
function similar to that for present day stars, in that the majority of the mass resides in
the lower-mass objects. This contrasts with the \zsix~and primordial clouds, in which the
bulk of the cluster mass is in high-mass stars.}

\end{figure*}

\section{Details of the Calculations} 

\subsection{Numerical Method} 
\label{code} 

We follow the evolution of the gas in this study using Smoothed Particle
Hydrodynamics (SPH). Our version of the code is essentially a parallelized version of
that used by Bate et~al.~(1995), which utilizes adaptive smoothing lengths for the gas
particles and a binary tree algorithm for computing gravitational forces and
constructing SPH neighbor lists. The calculations were performed on the J{\"u}lich
Multi-processor (JUMP) supercomputer at the John von Neumann Institute for Computing,
Research Center J{\"u}lich, Germany.

The thermal evolution of the gas in our simulations is modelled using a tabulated
equation of state (EOS) that is based on the results of the detailed chemical
modelling of \citet{om05}. We use their reported results for the temperature and
molecular fraction as functions of gas density (their Figures 1 \& 3) with
metallicities  ${\rm Z} = 0$, $10^{-6}$ and $10^{-5} \: {\rm Z_{\odot}}$ to compute 
the internal energy density and thermal pressure of the gas at a  range of different
densities. These values are used to construct look-up tables which are then used by
the SPH code to compute the pressure and internal energy at any required density, for
a given gas metallicity, via linear interpolation (in log-log space) between the tabulated values.
The resulting equations of state are plotted in Figure
\ref{fig:EOS}. By using a tabulated equation of state, we avoid the large
computational expense involved in solving the full thermal energy equation, while
still obtaining qualitatively correct behavior. 

To model the star formation in this study we use sink particles, as described by Bate
et~al.\ (1995). This involves replacing the innermost parts of dense, self-gravitating 
regions of gas with particles that can both accrete further material from their 
surroundings and interact with other particles in the simulation via gravity.  Sink
particles are formed once an SPH particle and its neighbors are gravitationally
bound, collapsing (negative velocity divergence), and within an accretion radius, 
$h_{acc}$, which is taken here to be 0.4 AU. Gravitational interactions between 
the sink particles and all other particles in the simulation are also smoothed, 
self-consistently, to $h_{acc}$. Our set-up allows us to identify sink particles 
as the direct progenitors of individual stars \cite[for a more detailed discussion, 
see e.g.][]{WK01}.

\subsection{Setup and Initial conditions}
\label{setup}

Our calculations are designed to start from the point where previous fragmentation 
calculations ended. It is now well established that the gas which falls into
collapsing dark matter mini-halos undergoes a phase of fragmentation, resulting in
the formation of large, self-gravitating clumps, with masses of order 100 \solmasp.
The origin of this fragmentation is the rapid cooling driven by ${\rm H_{2}}$ that
occurs at densities of around 10 -- $10^{4}$ \numdensu, and the masses of the clumps are
similar to the Jeans mass at the corresponding density and temperature in this
regime. We thus pick up the evolution from conditions similar to those reached in the
study of \citet[][]{bfcl01}. Since the process of chemical enrichment involves
supernovae events from the first stars we expect the gas to have a certain initial
level of turbulence, which is normally assumed to be absent during the formation of
the very first stars from purely primordial gas. We also adopt a net angular momentum
for the gas consistent with the results of simulations of cosmological structure
formation.

The clouds in this study have a mass of 500 \solmasp, and are modelled using either 2
million or 25 million SPH particles. In the higher resolution calculations, this gives a
particle mass of $2 \times 10^{-5} \: {\rm M_{\odot}}$ and a mass resolution of 0.002
\solmas (Bate \& Burkert 1997), roughly 10 times smaller than the opacity limit set by the
\citet[][]{om05} equation of state. These simulations therefore have roughly an order of
magnitude of surplus resolution. The low-resolution calculations have a mass resolution
roughly equal to the mass at which the gas becomes optically thick. To ensure that the Jeans
condition is not violated, the sinks are always formed just before the optically thick
regime in all the simulations. Our clouds have an initial radius of 0.17pc, at an initial
uniform density of $5 \times 10^{5}$~\numdensu. This corresponds to an initial free-fall
time of $t_{\rm ff} = (3\pi/32G\pi\rho)^{1/2} = 5.1 \times 10^{4}$ years. At this scale and
density regime, the contributions from dark matter to the gravitational potential are small
and are thus not taken into account in our computational set up. One can see from Figure
\ref{fig:EOS} that the different gas metallicities have slightly different internal energies
at the starting density. Thus, the \zleqsix~calculations have an initial ratio of thermal to
gravitational energy of $\alpha = E_{\rm therm}/ |E_{\rm grav}| =0.39$, while the cooler
\zfrag~calculations have an initial value of $\alpha = 0.32$. All simulations are given a
low level of initial turbulence, with the ratio of turbulent to gravitational potential
energy $E_{\rm turb} / |E_{\rm grav}| = 0.1$, and thus an RMS Mach number of
$\mathcal{M}_{RMS} \approx 1$. The clouds are set in initially uniform rotation, with $\beta
= E_{\rm rot}/ |E_{\rm grav}| = 0.02$. The conditions at the start of our simulation are thus
similar to those at which \cite{bcl02} form their sink particles (see their Figure 5 for
comparison).We perform low-resolution simulations for all three metallicity cases, and
high-resolution simulations for the primordial and \zfrag~cases.

\begin{figure*}[t]
\centerline{
\includegraphics[width=3.1in,height=3.1in]{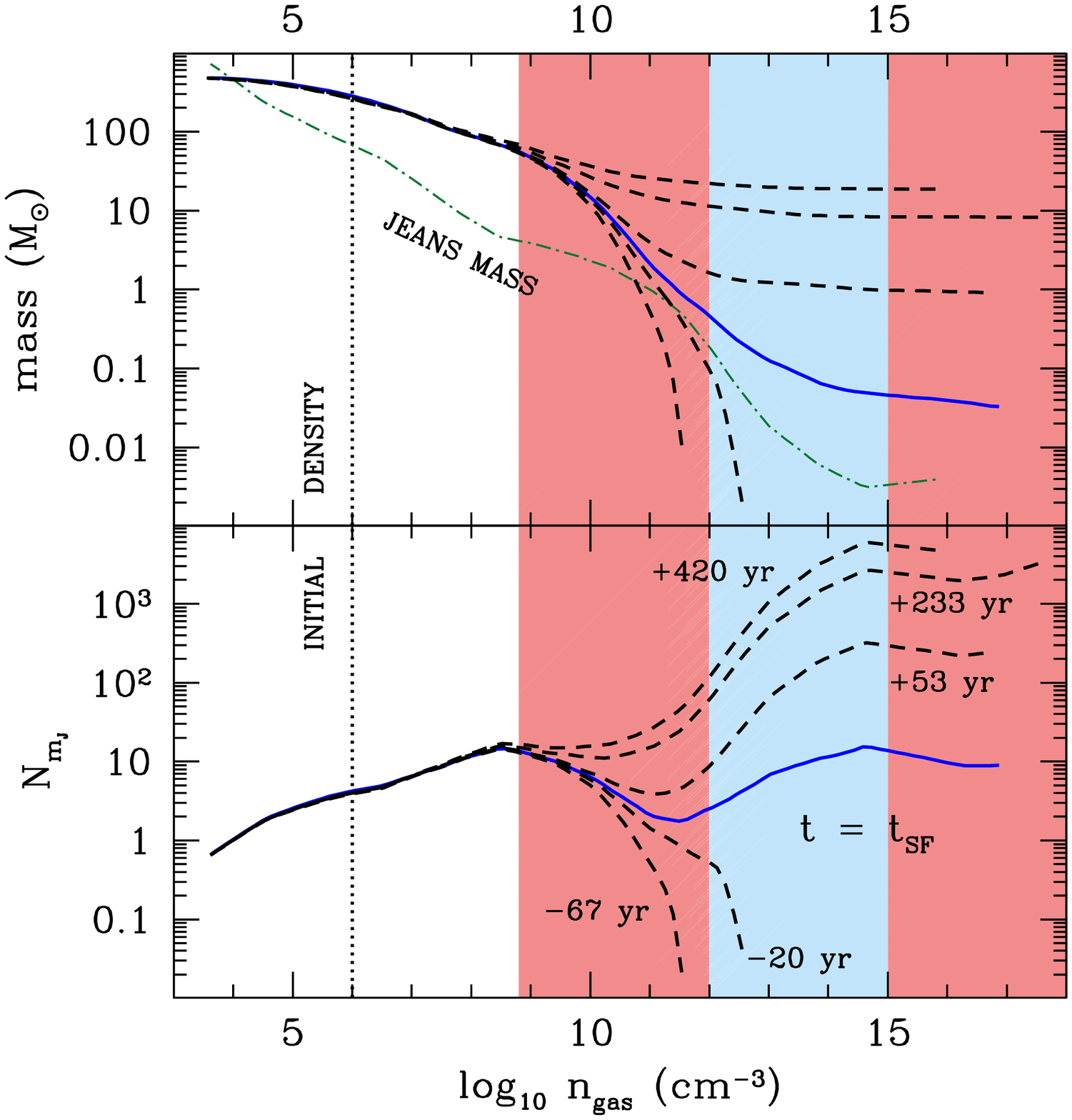}
\includegraphics[width=3.1in,height=3.1in]{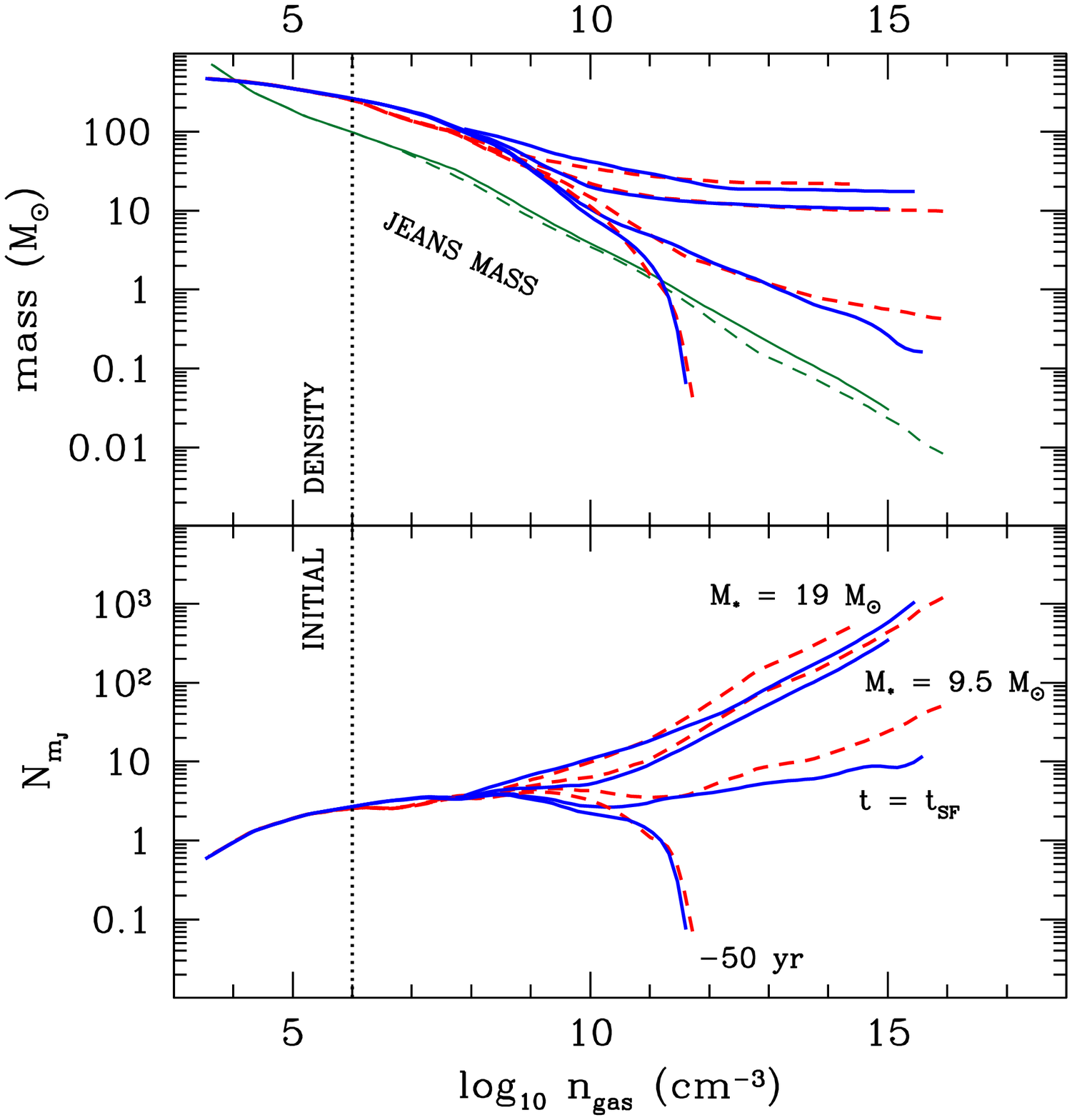}
}

\caption{ \label{fig:massrho} The mass (upper panels) and the number of Jeans masses (lower panels) are both
shown here as a function of gas density, with the high-resolution \zfrag~simulation in the
left-hand plot and both the high-resolution primordial and low-resolution \zsix~simulations
in the right-hand plot. In the upper panels, we plot the mass residing above a density,
$n_{\rm gas}$, as a function of that density, as well as the Jeans mass. In the lower
panels, we plot the number of Jeans masses as a function of density. In the left-hand panel,
we show via the shaded areas, the heating (pink) and cooling (light blue) phases in the
\zfrag~EOS. In the right-hand plot, we show the conditions in the primordial
(high-resolution) simulation (solid lines), and \zsix~simulation (dashed lines). The gas
conditions are shown at the point of star formation and 50 years earlier, as well as at two
instants after star formation, when the clouds have converted around 9.5\,\solmas and
19\,\solmas of gas into protostars. Note that we only label the evolutionary stages in the
lower panels, but the same progression with time, going from lower left-most lines to those
at the top, applies also to the upper panels. }
\end{figure*}

\section{Results}
\label{results}

We find that enrichment of the gas to a metallicity of only ${\rm Z} = 10^{-5} \:{\rm
Z_{\odot}}$ dramatically enhances fragmentation within the dense, collapsing cloud.
We first  focus our discussion on the evolution of the ${\rm Z} = 10^{-5} \: {\rm
Z_{\odot}}$ gas cloud, before contrasting with the  $10^{-6} \: {\rm Z_{\odot}}$ and
primordial clouds.

The evolution of the high-resolution \zfrag~simulation is illustrated in Figure
\ref{fig:sequence}, in a series of snapshots showing the density distribution of the
gas. We show several stages in the collapse process, spanning a time interval from
shortly before the formation of the first protostar (as identified by the formation of
a sink particle in the simulation) to 420 years afterwards. Note that we show only the
innermost 1\% of the full computational domain. A complementary view is given in Figure
\ref{fig:density}, which shows the particle densities as a function of position.

During the initial contraction, the cloud builds up a central core with a density of
about $n = 10^{10}\,$cm$^{-3}$. This core is supported by a combination of thermal
pressure and rotation. Eventually, the core reaches high enough densities to go into
free-fall collapse, and forms a single protostar. As more high angular momentum
material falls to the center, the core evolves into a disk-like structure with density
inhomogeneities caused by low levels of turbulence.  As it grows in mass, its density
increases. When dust-induced cooling sets in, it fragments heavily into a tightly
packed protostellar cluster within only a few hundred years. One can see this behavior
in particle density-position plots in Figure \ref{fig:density}. We stop the simulation
420 years after the formation of the first stellar object (sink particle). At this
point, the core has formed 177 stars. The evolution in the low-resolution simulation is
very similar. The time between the formation of the first and second protostars is
roughly 23 years, which is two orders of magnitude higher than the free-fall time at
the density where the sinks are formed. Without the inclusion of sink particles, we
would only have been able to capture the formation of the first collapsing object which
forms the first protostar: {\em the formation of the accompanying cluster would have
been missed entirely}.

The mass functions of the protostars at the end of the \zfrag~simulations (both high
and low resolution cases) are shown in Figure \ref{fig:masses} (left-hand panel). When
our simulation is terminated, the sink particles hold $\sim$ 19 \solmas of gas in
total. The mass function peaks somewhere below $0.1\ $M$_{\odot}$ and ranges from below
0.01$\,$M$_{\odot}$ to about $5\ $M$_{\odot}$. It is important to stress here that this
is not the final protostellar mass function. The continuing accretion of gas by the
cluster will alter the mass function, as will mergers between the newly-formed
protostars (which cannot be followed using our current sink particle implementation).
Protostellar feedback in the form of winds, jets and H{\sc ii} regions may also play a
role in determining the shape of the final stellar mass function. However, a key point
to note is that the chaotic evolution of a bound system such as this cluster ensures
that a wide spread of stellar masses will persist. Some stars will enjoy favourable
accretion at the expense of others that will be thrown out of the system (as can be
seen in Figure \ref{fig:sequence}), thus having their accretion effectively terminated
(see for example, the discussions in Bonnell \& Bate 2006 and Bonnell, Larson \&
Zinnecker 2007). The survival of some of the low mass  stars formed in the cluster is
therefore inevitable.

Our calculations demonstrate that the dust-cooling model of \citet{om05} can indeed
lead to the formation of low-mass objects from gas with very low metallicity. This
suggests that the transition from high-mass primordial stars to Population II stars
with a more ``normal'' mass spectrum occurs early in the universe, at metallicities at
or below Z $\approx10^{-5}\ $Z$_{\odot}$. Our simulations even predict the existence of
brown dwarfs in this metallicity range. Their discovery would be a critical test for
our model of the formation of the first star cluster. The first hints of its validity
come from the very low metallicity sub-giant stars that have recently been discovered
in the Galactic halo \cite[][]{christlieb02,bc05}, which have iron abundances less than
$10^{-5}$ times the solar value and masses below one solar mass, consistent with the
range reported here. 

Turning our attention to the ${\rm Z} = 0$ and \zsix~calculations, we also find that
fragmentation of the gas occurs, albeit at a much lower level than in the ${\rm Z} =
10^{-5} \: {\rm Z_{\odot}}$ run. The mass functions from these simulations are shown in
Figure \ref{fig:masses} (middle and right-hand panels), and are again taken when $\sim$
19 \solmas of gas has been accreted onto the sink particles, the same amount as is
accreted by the end of the \zfrag~calculations. 

The primordial gas clouds form fewer protostars than the \zfrag~clouds, with the high
resolution simulation forming 25 sink particles and the low-resolution simulation
forming 22 sink particles. The mass functions are considerably flatter than the present
day IMF, in agreement with the suggestion that Population III stars are typically very
massive. The fragmentation in the \zsix~simulation is slightly more efficient than in
the primordial case, with 33 objects forming. Again we stress that there is a delay of
several local free-fall times between the formation of first and second protostars in
these simulations: {\em without the inclusion of sink particles, we would have missed
the formation of the lower mass objects.}

\section{Conditions for Fragmentation and Cluster Formation}
\label{discussion}

\begin{figure}[htbp]
\begin{center}
\includegraphics[width=8.5cm]{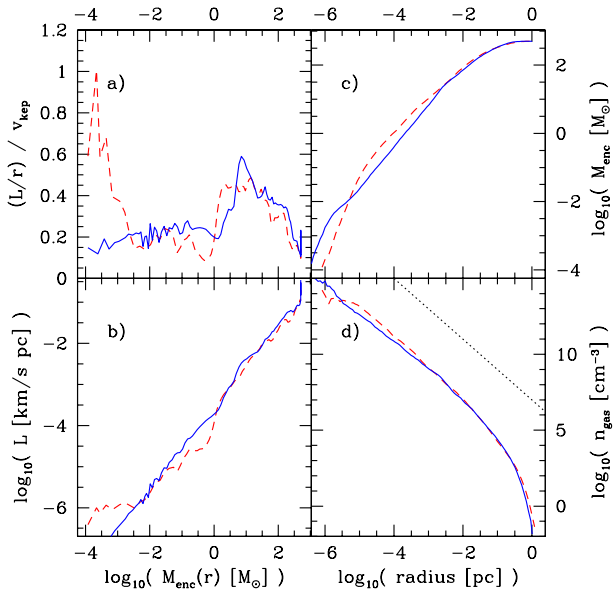}
\end{center}

\caption{\label{fig:ang-mom} $(a)$ Radially-averaged rotational velocity $L/r$ normalized
to the Keplerian velocity $v_{\rm kep} = (GM_{\rm enc}(r)/r)^{1/2}$ and $(b)$ absolute
value of the specific angular momentum $L$ in spherical shells centered on the density
peak as function of enclosed mass; $(c)$ $M_{\rm enc}(r)$ and $(d)$ average local gas
density $n_{\rm gas}(r)$ as function of distance $r$ to the center. Blue (solid line) 
curves denote gas
with  Z$=10^{-5}$Z$_{\odot}$ and red (dashed line) curves denote gas with Z$=0$. To guide your eye, a
power-law slope of $-2$ is indicated in $(d)$.} 

\end{figure}

We now discuss the physical origin of the fragmentation in the \zfrag~simulation, and
investigate the properties of the forming cluster. 
First, we focus on the distribution of gas in the center of the halo right at the onset of star formation, i.e.\ at the time when we identify the first sink particle. Figure \ref{fig:ang-mom} shows $(a)$ the distribution of rotational speed $L/r$ relative to the Keplerian velocity $v_{\rm kep} = (GM/r)^{1/2}$ and $(b)$ the specific angular momentum $L$ in spherical mass shells around  the halo center as a function of the enclosed mass. The blue curves denote the  $10^{-5}\,$Z$_{\odot}$ case, while the red-dashed curves illustrate the behavior of zero metallicity gas, as discussed further below.   It is evident that at the time when the first sink particle forms, rotational support plays  only a minor role.  The rotational velocity of gas in the inner parts of the halo is still sub-Keplerian by a factor or 4 to 5. This will change rapidly in the subsequent evolution as more and more higher angular momentum gas falls to the center.  In $(c)$ we plot the enclosed mass and in $(d)$ the spherically averaged density as a function of distance from the center. The density roughly follows a power-law distribution with slope $-2\pm0.1$.

We point out that the properties of the star forming clump  in our model  are virtually identical to those arising from full cosmological calculations taking into account the combined evolution of baryons and dark matter over time \cite[for comparison, see, e.g.][]{abn02,yoha06}. If anything, one could argue that the specific angular momentum we consider is somewhat on the low side. We therefore expect that fragmentation will  also occur in full cosmological simulations, if these are continued beyond the formation of the very first stellar object \cite[see e.g.\ ][for some preliminary evidence of this]{yokh07}.

To illustrate how the conditions
for cluster formation arise we show in Figure \ref{fig:density} the density
distribution perpendicular to the rotational axis of the system at different times.  In
Figure \ref{fig:massrho}, in the top panel, we plot both the mass of gas which resides
above a density $n$ as well as the Jeans mass. In the bottom panel, we plot
the number of Jeans masses, which is simply the mass at this density (or higher)
divided by the corresponding Jean mass.

\begin{figure*}[t]
\begin{center}
\unitlength1cm
\begin{picture}(20,5.5)
\put(1.0, 0.50){\includegraphics[width=1.8in,height=1.8in]{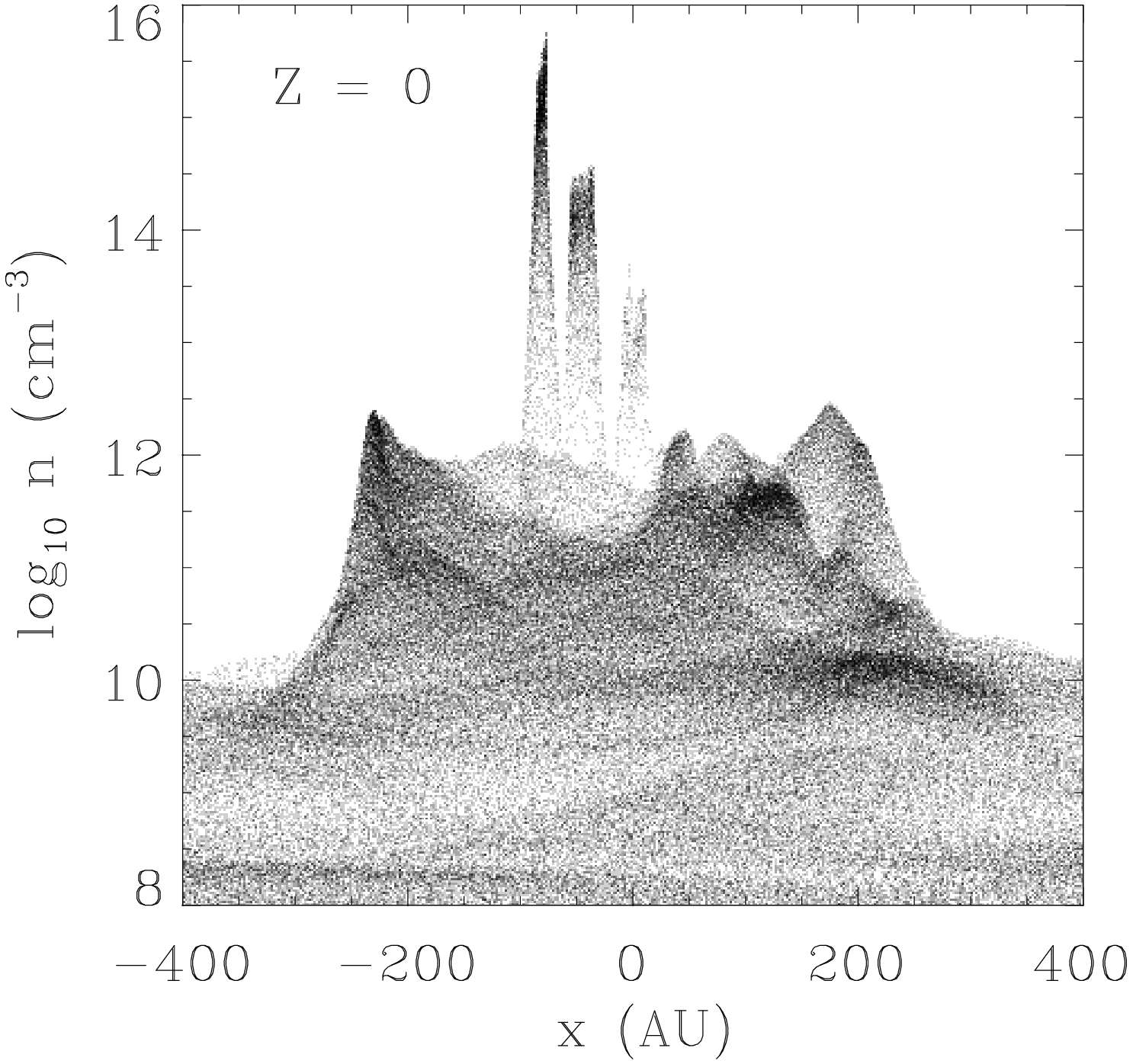}}
\put(7.0, 0.50){\includegraphics[width=1.8in,height=1.8in]{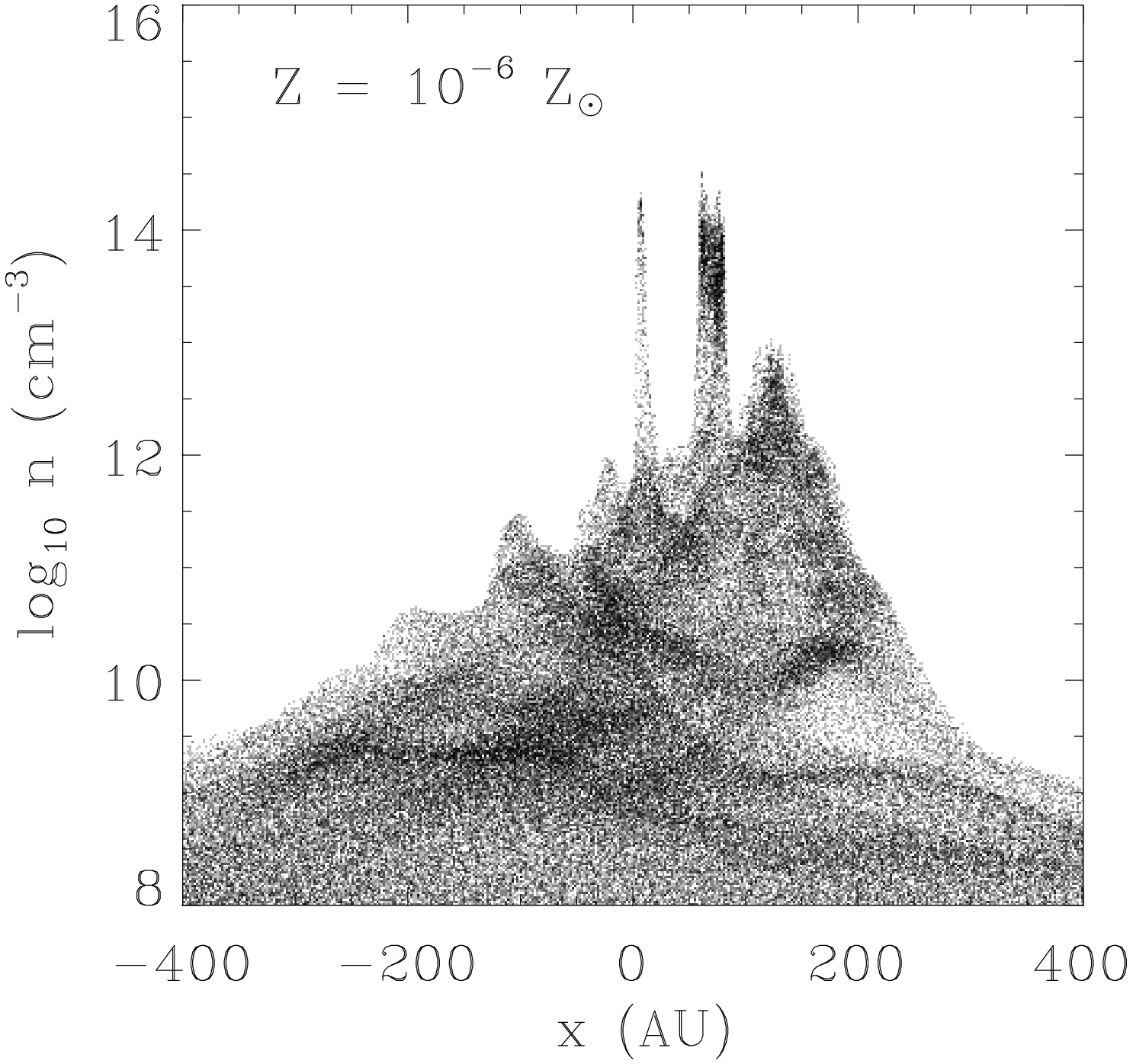}}
\put(13.0, 0.50){\includegraphics[width=1.8in,height=1.8in]{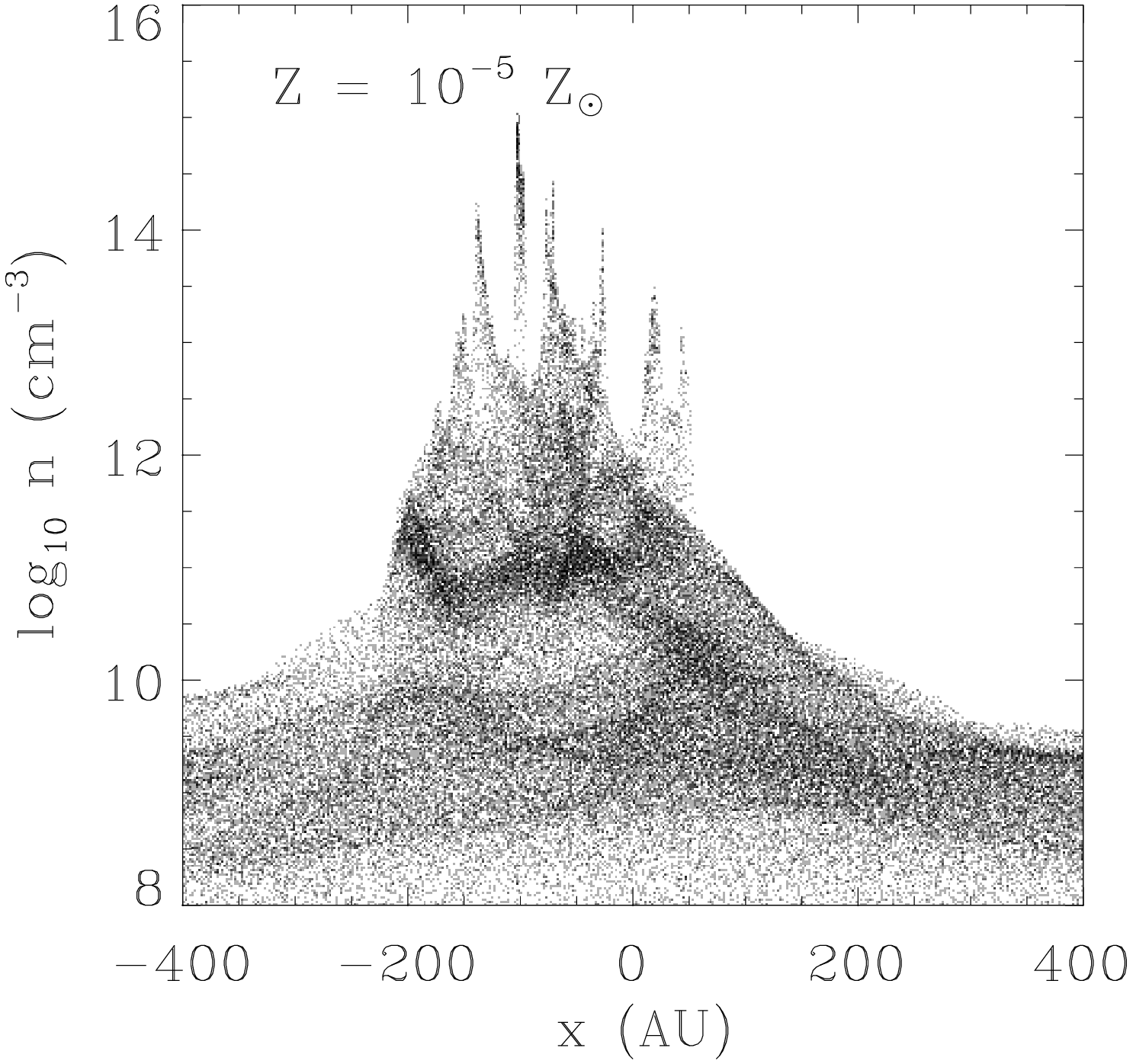}}
\end{picture}
\end{center}

\caption{\label{fig:lowresxrho} Particle densities as a function of position in the
low-resolution simulations, for the primordial (left), \zsix~(middle) and
\zfrag~simulations (right). The particles are plotted once the protostars in each
simulation have accreted 19 \solmas of gas.}

\end{figure*}

The fragmentation in the low-metallicity gas (the \zfrag~case) is the result of two 
key features in its thermal evolution. First, the rise in the EOS curve between
densities $10^{9}$cm$^{-3}$ and $10^{11}$cm$^{-3}$ causes material to loiter at this
point in the gravitational contraction. A similar behavior at densities around $n =
10^3\,$cm$^{-3}$ is discussed by Bromm et al.\ (2001).
The rotationally stabilized disk-like structure, as seen in the plateau at $n \approx
10^{10}$cm$^{-3}$ in Figure \ref{fig:density}, is able to accumulate a significant
amount of mass in this phase and only slowly increases in density. Second, once the
density exceeds $n \approx 10^{12}$cm$^{-3}$, the sudden drop in the EOS curve lowers
the critical mass for gravitational collapse by two orders of magnitude. The Jeans mass
in the gas at this stage is only $M_{\rm J} = 0.01\ $M$_{\odot}$, as visible in the top
panel of Figure \ref{fig:massrho}. The disk-like structure suddenly becomes highly
unstable against gravitational collapse. We see this when looking at the behavior of
the number of Jeans masses $N_{J}$  above a certain density $n$ in the bottom panel of
Figure \ref{fig:massrho}, which shortly after the formation of the first stellar object
increases to values as high as $N_{\rm J} \approx 10^3$. Consequently, the disk
fragments vigorously on timescales of several hundred years. A very dense cluster of
embedded low-mass protostars builds up, and the protostars grow in mass by accretion
from the available gas reservoir. The number of protostars formed by the end of our
simulation (177) is nearly two orders of magnitude larger than the initial number of
Jeans masses in the cloud set-up.

Because the evolutionary timescale of the system is extremely short -- the free-fall
time at a density  of $n = 10^{13}\,$cm$^{-3}$ is of the order of 10 years  --   none
of the protostars that have formed  by the time that we stop the simulation have yet
commenced hydrogen burning, justifying our decision to neglect the effects of
protostellar feedback in this study. Heating of the dust due to the significant
accretion luminosities of the newly-formed protostars will occur \citep{krum06}, but is
unlikely to be important, as the temperature of the dust at the onset of dust-induced
cooling is much higher than in a typical Galactic protostellar core ($T_{\rm dust} \sim
100 \: {\rm K}$ or more, compared to $\sim 10 \: {\rm K}$ in the Galactic case). The
rapid collapse and fragmentation of the gas also leaves no time for dynamo
amplification of magnetic fields \citep{tb04},  which in any case are expected to be
weak and dynamically unimportant in primordial and very low metallicity gas
\citep{wid02}.

The cluster forming in our  \zfrag~simulation  represents a very extreme analogue of
the clustered star formation that we know dominates in the present-day Universe
\cite[][]{ll03}. A mere 420 years after the formation of the first object, the
cluster has formed 177 stars (see Figure \ref{fig:sequence}). These occupy a region
of only around 400~AU, or $2 \times 10^{-3}$~pc, in size, roughly a hundredth of the
size of the initial cloud. With $\sim 19 \,$M$_{\odot}$  accreted at this stage, the
stellar density is $2.25 \times 10^{9}$ \solmas pc$^{-3}$. This is about five orders
of magnitude greater than the stellar density in the Trapezium cluster in Orion
\cite[][]{hh98} and about a thousand times greater than that in the core of 30
Doradus in the Large Magellanic Cloud \cite[][]{mh98}. This means that dynamical encounters 
will be extremely important during the formation of the first star cluster. The
violent environment causes stars to be thrown out of the denser regions of the cluster,
slowing down their accretion. The stellar mass spectrum thus depends on both the
details of the initial fragmentation process \cite[e.g.\ as discussed by][]{jappsen05,
clark05} as well as dynamical effects in the growing cluster \cite[][]{bcbp2001,bbv04}. 
This is different to present-day star formation, where the
situation is less clear-cut and the relative importance of these two processes may
vary strongly from region to region \cite[][]{kmk05, bb06, blz07}. 

Dynamical encounters in the extremely dense protocluster will also influence the binary
fraction of the stars that form. Wide binaries will be rapidly disrupted, and so any
binary systems that survive will be tightly-bound close binaries \cite[][]{kroupa98}.
Recent observations suggest that extremely metal-poor low-mass stars  have a higher
binary fraction than that found for normal metal-poor stars, and that the period
distribution of these binary systems is also skewed towards tight, short-period
binaries (Lucatello et~al.\ 2005). Mass transfer from a close binary companion may also
be able to explain the extremely high [C/Fe] ratios measured in the most metal-poor
stars currently known \citep{ry05,kom07}. Our results suggest that these stars may
originate in conditions similar to those that we find in our simulations. Further work
aimed at improving our understanding of the binary statistics of low-metallicity stars
formed by dust-induced fragmentation is clearly required.

As mentioned in Section \ref{results}, the primordial and the \zsix~cases also
exhibit some fragmentation. Careful analysis of the \citet{om05} EOS for
zero-metallicity gas shows roughly isothermal behavior in the density range
$10^{14}\,$cm$^{-3} \le n \le 10^{16}\,$cm$^{-3}$, i.e.\ just before the gas becomes
optically thick and begins to heat up adiabatically. Conservation of angular momentum
again leads to the build-up of a rotationally supported massive disk-like structure
which then fragments into several objects. This is understandable, as isothermal disks
are susceptible to gravitational instability \cite[][]{bodenheimer95} once they have
accumulated sufficient mass. Further, Goodwin et~al.~(2004a,b) show how even very low
levels of turbulence can induce fragmentation. Since turbulence creates
local anisotropies in the angular momentum on all scales, it can always provide some 
centrifugal support against gravitational collapse. This support can then provide a window in 
which fragmentation can occur.

In addition to the quasi-isothermal behavior of the gas, both of the
\zleqsix~equations of state from Omukai et~al. (2005) also contain a brief phase of
cooling during the  collapse, which further aids fragmentation. In the primordial case,
this occurs very late in the collapse, just above $n = 10^{14}$ \numdensu, and is due
to the onset of  efficient cooling from ${\rm H_{2}}$ collision-induced emission
\citep{on98,ra04}. In the \zsix~EOS, the cooling occurs earlier, at around $n = 10^{10}$
\numdensu, and is due to a combination of effects -- enhanced ${\rm H_{2}}$ cooling
resulting from the rapid increase in the molecular fraction at these densities due to
efficient three-body  ${\rm H_{2}}$ formation, and rotational and vibrational line
cooling from ${\rm H_{2}O}$ -- that are present in the model of Omukai et~al.\ (2005). 
One can see  the emergence of structure at these densities
in the particle plots shown in Figure~\ref{fig:lowresxrho}. However, the effect is much less
pronounced in the primordial case. Indeed a further low-resolution simulation of the
primordial EOS in which this dip was removed yielded almost identical results, forming
17 sink particles instead of 22, suggesting that the quasi-isothermal nature of the 
gas is more important than this brief cooling phase.

In comparison to the \zfrag~case, the strength of fragmentation in the
\zleqsix~calculations is weak, and only a few objects form for the combination of total
mass ($M=500\,$M$_{\odot}$), angular momentum ($E_{\rm rot}/ |E_{\rm grav}|= 0.02$),
and level of initial turbulence ($E_{\rm turb}/|E_{\rm grav}| = 0.1$) that we adopted
in our simulations. Consequently the stars (sink particles) forming in the
\zleqsix~simulations are typically of higher mass than those in the \zfrag~simulations,
consistent with the predictions made by \citet[][]{abn02} and \citet[][]{bcl02}.
Although the recent high-resolution SPH simulation of primordial gas performed by
\citet{yoha06}, which predict a similar EOS to that used in our primordial simulations,
find no fragmentation, they do not follow the evolution of the gas beyond the formation
of the first collapsing core, since they do not include  sink particles in their study,
and so they may miss this subsequent phase of  fragmentation. However, more
recent work \citet{yokh07}, does show evidence of fragmentation on scales of around
0.1pc.  

\section{Caveats} 
\label{caveats} 

Although the equations of state taken from Omukai et~al.\ (2005) provide an opportunity
for  vigorous fragmentation at low metallicities, there are a number of questions which
still need to be addressed. One immediate uncertainty is the applicability of the
results derived from a 1-zone model to a full three-dimensional collapse calculation,
which contains  turbulence and thus local anisotropies in velocity and density
structure. In particular,  the existence of a strict relationship between temperature
and density is unlikely, even when the cooling time of the gas is short compared to the
dynamical time (Whitehouse \&  Bate 2006). If the gas were to  collapse more slowly
than is assumed in the Omukai et~al.\  (2005) calculations, or if the opacity of the
gas were to be lower, then less heating would occur prior to the onset of efficient
dust cooling. 

On the other hand, if the  collapse is faster than Omukai et~al.\ (2005) assume or if
the gas opacity is greater, then more heating would occur. Since the amount of heating
that occurs during the loitering phase helps to accumulate material at high densities,
which then acts as a reservoir for fragmentation, the amount of fragmentation will be
sensitive to the thermal evolution of the gas during this phase of the collapse, and is
therefore somewhat uncertain. However, we note that unless the temperature dip is
eliminated entirely (which seems unlikely), fragmentation will still occur along the
lines outlined in this paper. We therefore believe our results to be qualitatively, if
not quantitatively, correct.

A related issue is one of dust opacity. In their study, Omukai et~al.\ (2005) use a
dust model based on Pollack et~al.\ (1994), which was designed for the study of
Galactic dust. However, it is not at all clear that this is the appropriate model  to
use, given that high-redshift dust is likely to differ significantly from local dust
\citep[see e.g.][]{tod01}. Schneider et~al.\ (2006) have performed a similar  study to
that of Omukai et~al.\ (2005), but use a dust model based on the results of
\citet{tod01} and \citet{sfs04} in place of the Pollack et~al.\ (1994) dust model. 
They find that this leads to qualitatively similar behavior to the Omukai et~al.
(2005)  model, but that the onset of effective dust cooling happens at lower
metallicity,  \zsix~rather than \zfrag. However, the use of this alternative dust model
can also be  questioned, since \citet{noz03} have shown that the composition and
properties of the  dust produced by population III supernovae are strongly dependent on
the (poorly-constrained)  degree of mixing assumed to occur in the ejecta, and since
the model does not take into  account the destruction of grains by thermal sputtering
in the reverse shock in the supernova remnant \citep{bs07}, or the growth of grains
through accretion or coagulation  during the protostellar collapse \citep[see
e.g.][]{fpw05}.

Lastly, the total amount of heating produced by three body H$_{2}$ formation is
uncertain. This reaction, which takes place primarily at gas number densities  between
$10^{8}$ and $10^{13} \: {\rm cm^{-3}}$, is responsible for much of the  rise in gas
temperature prior to the dust-cooling phase. However, there is a  difference of two
orders of magnitude between the three-body ${\rm H_{2}}$ formation rate used in the
simulations of primordial star-formation performed by  Abel et~al.~(2002) and the most
recent theoretical determination by Flower \&  Harris (2007), with other suggested
rates spanning the full range in between.  This introduces an additional uncertainty
into the thermal evolution of the gas during the high-density loitering phase.

\section{Conclusions}
\label{conclusions}

We have performed numerical simulations of star formation in very high density gas (in
the range $5 \times 10^{5} \leq n \leq  10^{16} \: {\rm cm^{-3}}$) in the early
universe. The aim of the study was to investigate whether the dust-induced
fragmentation predicted by Omukai et~al.\ (2005) and Schneider et~al.\ (2002, 2006) 
does actually occur in realistic systems, and to begin to constrain the resulting IMF. 
The major differences of our work from the only previous numerical study of
dust-induced  fragmentation (Tsuribe \& Omukai 2006) are  the inclusion of the effects
of rotation,  which prove to be of vital importance, and the use of sink particles to
capture the  formation of multiple protostellar objects.

Based on the equations of state reported by Omukai et~al.\ (2005), our results show
that fragmentation of protogalactic gas at very high densities above $n_{\rm gas} =
10^{12}\,$cm$^{-3}$ is almost unavoidable, as long as the angular momentum is
non-negligible. In this case, rotation leads to the build-up of a massive disk-like
structure which provides the background for smaller-scale density fluctuations to grow,
some of which become gravitationally unstable and collapse to form stars. At
metallicities above Z~$\sim 10^{-5}\,$Z$_{\odot}$ dust cooling becomes effective at
densities $n_{\rm gas} \sim 10^{12}\,$cm$^{-3}$ and leads to a sudden drop of
temperature which in turn induces vigorous fragmentation. A very dense cluster of
low-mass protostars builds up, which we refer to as the first stellar cluster. The mass
spectrum peaks below $1\,$M$_{\odot}$, which is similar to the value in the  solar
neighborhood \cite[][]{kroupa02,chabrier03} and is also comparable to the mass of the
very low metallicity subgiant stars recently discovered in the halo of our Milky Way
\citep{christlieb02, bc05}. If the dust induced cooling model proposed by Omukai et al
(2005) is accurate, then the high-density, low-metallicity, fragmentation we describe
here may be the dominant process which shapes the stellar mass function.

We find that even  purely primordial Z~$ = 0$~gas with sufficient rotation may fragment
at densities $10^{14}\,$cm$^{-3} \le n \le 10^{16}\,$cm$^{-3}$. In this density range,
zero metallicity gas is roughly isothermal and the disk-like structure that forms due
to angular momentum conservation is marginally unstable. It  fragments into several,
quite massive objects, thus supporting the hypothesis that metal-free stars should have
masses in excess of several tens of M$_{\odot}$. Similar behavior is found for gas with
a metallicity of \zsix.

\begin{acknowledgments}

The authors would like to thank Kazuyuki Omukai, Tom Abel, Volker Bromm, Robi
Banerjee and Mordecai-Mark Mac Low for  helpful discussions concerning this topic.
In particular, we would also like to thank Naoki Yoshida for many interesting
and insightful discussions regarding this paper. All computations described here
were performed  at the J{\"u}lich Multi-processor (JUMP) supercomputer at the John
von Neumann Institute for Computing, Research Centre J{\"u}lich, Germany. PCC
acknowledges  support by the Deutsche Forschungsgemeinschaft (DFG) under grant
KL~1358/5.   SCOG acknowledges travel support from the European Commission FP6 Marie
Curie  RTN CONSTELLATION (MRTN-CT-2006-035890).

\end{acknowledgments}

\end{document}